\def\ts     {\thinspace}
\def\kms    {\ifmmode{{\rm \ts km\ts s}^{-1}}\else{\ts km\ts s$^{-1}$}\fi}
\def\msol   {\ifmmode{{\rm M}_{\odot} }\else{M$_{\odot}$}\fi}
\def\lsol   {\ifmmode{L_{\odot}}\else{$L_{\odot}$}\fi}
\def\lfir   {\ifmmode{L_{\rm FIR}}\else{$L_{\rm FIR}$}\fi}
\def\zsol   {\ifmmode{{\rm Z}_{\odot}}\else{Z$_{\odot}$}\fi}

\def\aco    {\ifmmode{{\rm CO}(J\!=\!1\! \to \!0)}\else{{\rm CO}($J$=1$\to$0)}\fi}
\def\taco    {\ifmmode{^{12}{\rm CO}(J\!=\!1\! \to \!0)}\else{$^{12}${\rm CO}($J$=1$\to$0)}\fi}
\def\bco    {\ifmmode{{\rm CO}(J\!=\!2\! \to \!1)}\else{{\rm CO}($J$=2$\to$1)}\fi}
\def\cco    {\ifmmode{{\rm CO}(J\!=\!3\! \to \!2)}\else{{\rm CO}($J$=3$\to$2)}\fi}
\def\dco    {\ifmmode{{\rm CO}(J\!=\!4\! \to \!3)}\else{{\rm CO}($J$=4$\to$3)}\fi}
\def\eco    {\ifmmode{{\rm CO}(J\!=\!5\! \to \!4)}\else{{\rm CO}($J$=5$\to$4)}\fi}
\def\fco    {\ifmmode{{\rm CO}(J\!=\!6\! \to \!5)}\else{{\rm CO}($J$=6$\to$5)}\fi}
\def\gco    {\ifmmode{{\rm CO}(J\!=\!7\! \to \!6)}\else{{\rm CO}($J$=7$\to$6)}\fi}
\def\hco    {\ifmmode{{\rm CO}(J\!=\!8\! \to \!7)}\else{{\rm CO}($J$=8$\to$7)}\fi}
\def\ico    {\ifmmode{{\rm CO}(J\!=\!9\! \to \!8)}\else{{\rm CO}($J$=9$\to$8)}\fi}
\def\jco    {\ifmmode{{\rm CO}(J\!=\!10\! \to \!9)}\else{{\rm CO}($J$=10$\to$9)}\fi}
\def\kco    {\ifmmode{{\rm CO}(J\!=\!11\! \to \!10)}\else{{\rm CO}($J$=11$\to$10)}\fi}
\def\ci     {\ifmmode{{\rm C}{\rm \small I}}\else{C\ts {\scriptsize I}}\fi}
\def\hi     {\ifmmode{{\rm H}{\rm \small I}}\else{H\ts {\scriptsize I}}\fi}
\def\hh     {\ifmmode{{\rm H}_2}\else{H$_2$}\fi}
\def\cone {\ifmmode{{\rm C}{\rm \small I}(^3\!P_1\!\to^3\!P_0)}
    \else{C\ts {\scriptsize I}{\small$(^3\!P_1\!\to^3\!\!\!P_0)$}}\fi}
\def\ctwo {\ifmmode{{\rm C}{\rm \small I}(^3\!P_2\!\to^3\!P_1)}
     \else{C\ts {\scriptsize I}{\small$(^3\!P_2\!\to^3\!\!\!P_1)$}}\fi}
\def\cij {\ifmmode{{\rm C}{\rm \small I}\,(^3P_i\to^3P_j)}\else{C\ts {\scriptsize I}\,{\small$(^3P_i\to^3P_j)$}}\fi}
\def\cii    {\ifmmode{{\rm C}{\rm \small II}}\else{C\ts {\scriptsize II}}\fi}
\def\tex {\ifmmode{{T}_{\rm ex}}\else{$T_{\rm ex}$}\fi}
\def\tmb {\ifmmode{{T}_{\rm mb}}\else{$T_{\rm mb}$}\fi}
\def\tkin {\ifmmode{{T}_{\rm kin}}\else{$T_{\rm kin}$}\fi}
\def\microns {\ifmmode{\mu{\rm m}}\else{$\mu$m}\fi}
\def\nhh   {\ifmmode{n({\rm H}_2)}\else{$n$(H$_2$)}\fi}
\def\gradv {\ifmmode{{\rm dv/dr}}\else{dv/dr}\fi}
\documentclass[traditabstract]{aa} % for the abstract without structuration
\usepackage[varg]{txfonts}
\usepackage{natbib,amssymb}
\usepackage{graphicx}
\usepackage{multirow}
\bibpunct{(}{)}{;}{a}{}{,}
\begin{document}
\title{GREAT confirms transient nature of the circumnuclear disk}
\author{M.A. Requena-Torres\inst{\ref{inst1}}
	  \and
      R. G\"usten\inst{\ref{inst1}}
      \and
       A. Wei\ss\inst{\ref{inst1}}
      \and
	  A.I. Harris\inst{\ref{inst2}}
	  \and
	  J. Mart\'in-Pintado\inst{\ref{inst3}}
      \and
	  J. Stutzki\inst{\ref{inst4}}
      \and
\\
	  B. Klein\inst{\ref{inst1}}
	  \and
	  S. Heyminck\inst{\ref{inst1}}
          \and
	  C. Risacher\inst{\ref{inst1}}
	  }

   \institute{Max-Planck-Institut f\"ur Radioastronomie, Auf dem H\"ugel 69, 53121 Bonn, Germany\\
              \email{mrequena@mpifr-bonn.mpg.de} \label{inst1}
        \and
         Department of Astronomy, University of Maryland, College Park, MD 20742, USA\label{inst2}
        \and
         Centro de Astrobiolog\'ia (CSIC-INTA), Ctra. deTorrej\'on Ajalvir,  E-28850 Torrej\'on de Ardoz, Madrid, Spain \label{inst3}
               \and
I. Physikalisches Institut der Universit\"at zu K\"oln, Z\"ulpicher Stra§e 77, 50937 K\"oln, Germany\label{inst4}
        }

   \date{Received February 17, 2012; accepted March 27, 2012}

   \abstract{
    We report SOFIA/GREAT, {\it Herschel}/HIFI, and
     ground-based velocity-resolved spectroscopy of carbon monoxide
     (CO) rotational transitions from J=2-1 to J=16-15 toward two
     positions in the circum-nuclear disk (CND) in our Galactic
     center.  Radiative transfer models were used to derive information
     on the physical state of the gas traced by CO.
     The excitation of the CO gas cannot be explained by a single
     physical component, but is clearly the superposition of various
     warm gas phases.  In a two-component approach, our large 
     velocity gradient (LVG) analysis
     suggests high temperatures of $\sim$200 K with moderate gas
     densities of only $\sim$10$^{4.5}$ cm$^{-3}$ for the bulk of the
     material. A higher excited phase, carrying $\sim$20-30$\%$ of the
     column densities, is warmer ($\sim$300-500 K) but only slightly
     denser ($\sim$10$^{5.3}$ cm$^{-3}$).  These densities are too low
     to self-stabilize the clumps against their high internal
     turbulence and fall below the Roche density ($>$10$^7$ cm$^{-3}$)
     at 1.5 pc galactocentric distance.  We conclude that the bulk of
     the material in the CND is not organized by self-gravity nor
     stable against tidal disruption, and must be transient.
}
  \keywords{ISM: clouds - ISM: kinematics and dynamics - ISM:
     molecules - Galaxy: center - Radio lines: ISM}

   \maketitle

%
%________________________________________________________________

\section{Introduction}

At a distance of only 8.5 kpc, our Galactic center (GC) is a unique
laboratory for studying the physical processes that
also occur within extragalactic nuclei in
general.  The innermost central parsecs contain all the ingredients
and processes that also affect the gas in other nuclei that are too
distant for detailed study.  Here we are most concerned with the
massive gas clouds that lie within a few parsecs of the Galactic
center's black hole, SgrA$^*$.  It is only in our own Galaxy's center that we
can spatially resolve and study these likely reservoirs of matter that
fuel the center's episodic activity.

The global distribution of the interstellar matter in the very center
of the Galactic nucleus is a  $\sim$1-1.5 pc radius
\emph{ionized cavity} centered on SgrA$^*$.  The cavity's outer
boundary forms the sharp inner edge of the \emph{circum-nuclear
  disk} (CND), a thin disk-like structure composed of dense molecular
gas in filaments and streamers that extend to radius $\sim$5 pc.
Ionized material is present also in the cavity in the form of streamers
with a ``mini-spiral'' shape.  These streamers have high radial
velocities and carry material deep into the central cavity and close
to SgrA$^*$ \citep[for more detailed references we refer to the review
by ][]{genzel2010}.

Despite many investigations, the physical characteristics
of the neutral gas remains a topic of debate.  Estimates of the CND's
molecular gas mass range between 10$^{4}$ (from dust) and 10$^{5-6}$
M$_\odot$ (derived from gas tracers).  Depending on the probe, density
estimates for the dense clumps are 10$^{5-8}$ cm$^{-3}$ and gas
temperatures run from 50 to a few hundred K.  The range of physical parameters
leaves open many questions about the nature and fate of the CND,
for instance whether the clumps' densities exceed the Roche limit and
make them stable against tidal disruption, or whether they are fully
transient features.

A prerequisite for a quantitative comparison between these competing
scenarios is an in-depth investigation of the temperature and density
distribution in this clumpy, filamentary ring.  Early work from Genzel
et al.\ (1985) detecting CO(16-15) with the KAO, clearly indicated the
existence of a gas phase with high temperatures.  \citet{har85} tied
low-, mid-, and high-J observations together for a full CO excitation
analysis that gave T$_{kin}$ $\sim$300 K and n(H$_2$)
$\sim$10$^{4.5}$ cm$^{-3}$. \citet{brad05}, revisiting these data with new CO(7-6) observations derived
warm 200--300 K and moderate densities 5--7$\times$10$^4$. \citet{whi03} used infrared ISO/LWS data to
find gas with T$_{kin}$ $\sim$900 - 1400 K and n(H$_2$) $\sim$ 10$^4$
- 10$^{4.5}$ cm$^{-3}$.

In this contribution we present initial results of a systematic and comprehensive
investigation of the CND, combining velocity-resolved heterodyne spectroscopic
observations with the far-infrared spectrometers GREAT\footnote{GREAT, the {\underline{G}}erman {\underline{RE}}ceiver for
{\underline{A}}stronomy at {\underline{T}}erahertz frequencies, is a development by the MPI f\"ur Radioastronomie and the KOSMA\,$/$\,Universit\"at zu K\"oln, in cooperation with the MPI f\"ur Sonnensystemforschung and the DLR Institut f\"ur Planetenforschung.} on board of
SOFIA and HIFI/PACS on {\it Herschel}, supplemented by data of lower-J
CO transitions from APEX and the IRAM-30m telescope.

Here we perform an in-depth CO excitation analysis toward the
prominent northern and southern lobes of the CND (Fig.~1).  The CO
spectral energy distribution is from line-resolved observations of
most transitions of the CO rotational ladder up to J = 16-15, and
additional $^{13}$CO isotopologues to confine the optical depth of the
CO lines.

\begin{figure}
\centering
  \includegraphics[width=8.5cm]{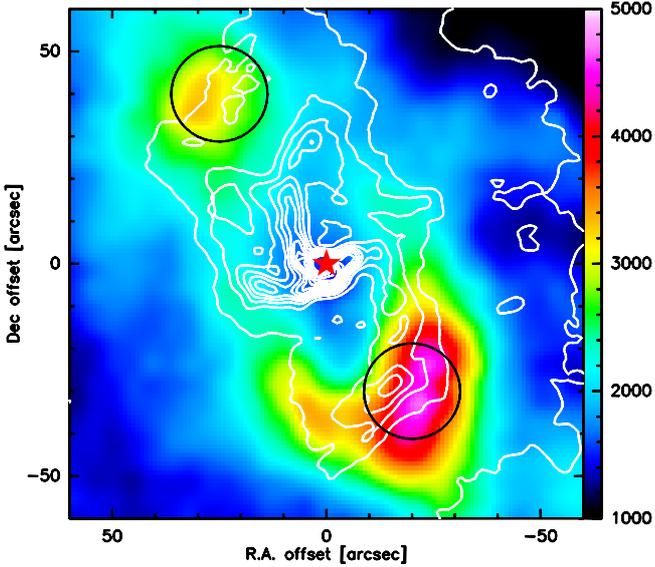}
  \caption{CND in the emission of CO(6-5), observed with the
    CHAMP$^+$ receiver at APEX. The CO intensity (color map in 
    K km/s, in T$_{mb}$) was integrated between -150 and 150 km/s. 
    The angular resolution of the map is 9.5$''$. VLA 6cm
    radio continuum \citep[contours, ][]{yusef87} shows ionized gas in 
    the "mini-spiral" near SgrA$^*$. The map is centered on SgrA$^*$ (red star), (EQ J2000)
    17$^o$45$'$39.9$''$, -29$^h$00$^m$28.1$^s$.  The two circles
    denote the southern (CND-S) and northern (CND-N) emission maxima
    toward which our excitation studies are performed, at offsets of
    (-20$''$,-30$''$) and (25$''$,40$''$). The diameter of the
    circles corresponds to the beam of GREAT at the frequency of
    the CO(11-10) transition (22.5$''$).}
    \label{co65map}
    \vspace{-0.5cm}
\end{figure}

\section{Observations}
The GREAT observations were performed during two SOFIA Basic Science flights in 2011. On April 08, GREAT was
operated at 42700 ft altitude in its L1a/L2 configuration \citep{hey12}, tuned to the CO(11-10) and CO(16-15) lines. In the second
flight, on July 16, $^{(13)}$CO(13-12), and CO(16-15) were observed in the L1b/L2 channels (40000 ft).  We used the Fast Fourier
Transform spectrometers \citep[FFTs, ][]{klein12} for back-ends; each provided 212 kHz spectral resolution across a bandwidth of 1.5 GHz.
At the high observing frequencies, GREAT's bandwidth is insufficient to cover all velocities in the Galactic center
in one observation.  A single setup does cover velocities from the individual positions, however, and we made two separate receiver
tunings offset by +75 and -75 km/s relative to rest to cover each position.

The observations were performed in double-beam switching mode, with a
throw of 360$''$ at a position angle of 45$^\circ$ (CCW to R.A.),
at a rate of 1 Hz.  We made 5$\times$3 and 3$\times$3 raster maps on
an 8$''$ grid in equatorial coordinates for CND-S and
CND-N, respectively.  Optical
guide cameras established pointing to an accuracy of $5^{\prime\prime}$.  The beam width $\Theta_{mb}$ is
22.5$''$ at the frequency of CO(11-10), 19.1$''$ for CO(13-12) and 15.6$''$ for
CO(16-15). The main beam and forward efficiencies are $\eta_{mb}$
=0.54 (L1) and 0.51 (L2), with $\eta_{f}$ =0.95.
The data were calibrated with the KOSMA/GREAT calibrator \citep{guan12}, carefully removing residual telluric lines, and
further processed with the GILDAS\footnote{http://www.iram.fr/IRAMFR/GILDAS} packages CLASS and GREG.

\begin{table}
\caption{Observed integrated line intensities}          % title of Table
\label{obsdata}      % is used to refer this table in the text
\centering                                                % used for centering table
\begin{tabular}{l c c c c c c c}        % centered columns (4 columns)
\hline\hline                              % inserts double horizontal lines
\multirow{2}{*}{Transition} &  \multirow{2}{*}{E$_{\rm{up}}${\tiny{[K]}}} &  \multirow{2}{*}{$\nu${\tiny{[THz]}}} &   \multicolumn{2}{c}{$\int$T$_{mb}\cdot$d$v$ {\tiny{[K km/s]}}}             &                        \\
           &  & &             \tiny{CND-S}        & \tiny{CND-N}     &   \\
\hline    \\[-2ex]
$^{12}$CO(2-1)   & 5.5  & 0.230  &  797.4           & 670.4    & I     \\
$^{13}$CO(2-1)   & 5.3  & 0.220  &   49.0           & 21.2     & I     \\
$^{12}$CO(3-2)   & 16.6 & 0.346  & 1668.3           & 852.9    & A     \\
$^{12}$CO(4-3)   & 33.2 & 0.461  & 2199.6           &1023.5    & A     \\
$^{12}$CO(6-5)   & 83.0 & 0.691  & 1782.3           & 786.7    & A     \\
$^{13}$CO(6-5)   & 79.3 & 0.661 &  165.5           &  67.2    & A     \\
$^{12}$CO(7-6)   &116.2 & 0.807  & 1753.0           & 708.6    & A     \\
\hline \\[-2ex]
$^{12}$CO(10-9)  &248.9 & 1.152  &  841.5           & 242.1    & H     \\
\hline \\[-2ex]
$^{12}$CO(11-10) &304.2 & 1.267  &  818.8           & 218.0        & G     \\
$^{12}$CO(13-12) &431.3 & 1.497 &  391.6           & 77.0        & G     \\
$^{13}$CO(13-12) &412.3 & 1.431  &  26.5           &         & G     \\
$^{12}$CO(16-15) &663.4 & 1.841  &  188.2           & 34.4        & G     \\
\hline                                                      % inserts single horizontal line
\hline                                                      % inserts single line
\end{tabular}
\tablefoot{The integration was performed in the velocity ranges [-150,-30] and [80,150] for CND-S and CND-N, respectively. All data were convolved to the 22.5$''$ beam of the GREAT CO(11-10) observations. For each transition we also give the rest frequency and energy of the upper level above ground is. Data are from observations with $\underline{\rm{A}}$PEX, the $\underline{\rm{I}}$RAM-30m telescope, $\underline{\rm{H}}$IFI/{\it Herschel} and $\underline{\rm{G}}$REAT/SOFIA.}
\vspace{-0.5cm}
\end{table}

To confine the CO excitation of the CND we have
mapped during recent years all CO mm/submm transitions, 2 $\leq$
J$_{up}$ $\leq$7 with APEX
and the IRAM-30m telescope,that are accessible from these large ground-based
facilities. These data will be presented elsewhere (Requena-Torres et al., in prep.);
for this study we extracted the spectra necessary for a  comparison with the
new SOFIA/GREAT FIR observations. We also included a CO(10-9) data point
from the HexGal key program on the Galactic center (PI: R.G\"usten).  Table \ref{obsdata} gives transition information and
velocity-integrated line intensities for the two positions in the CND
we analyze here, CND-S and CND-N (see Fig.~1).
All spectra were convolved from our Nyquist-sampled maps to the same angular resolution
(22.5$''$, of the CO(11-10) observations), thereby defining a uniquely
coherent data base.
\section{Results and analysis}
\begin{figure}[!tp]
\centering
\includegraphics[width=10.2cm]{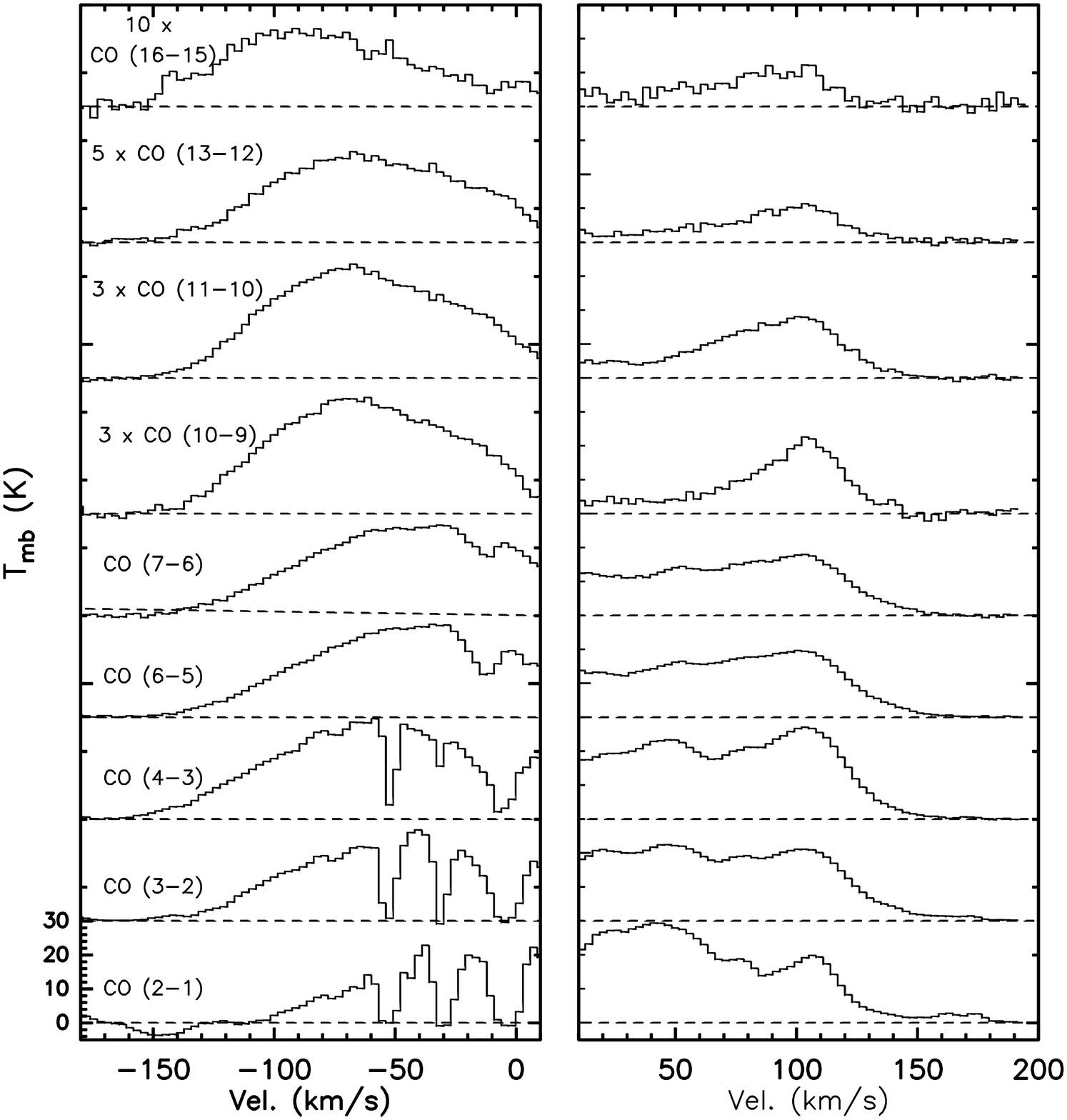}
\caption{CO spectra toward the southern (left) and northern (right) lobe of the CND. We display the relevant velocity ranges only. The data were box-smoothed to 3 km/s spectral resolution. All spectra are convolved to the same angular resolution of 22.5$''$.}
\label{cospec1}
\vspace{-0.5cm}
\end{figure}

\begin{figure}[!tp]
\centering
\includegraphics[width=9cm]{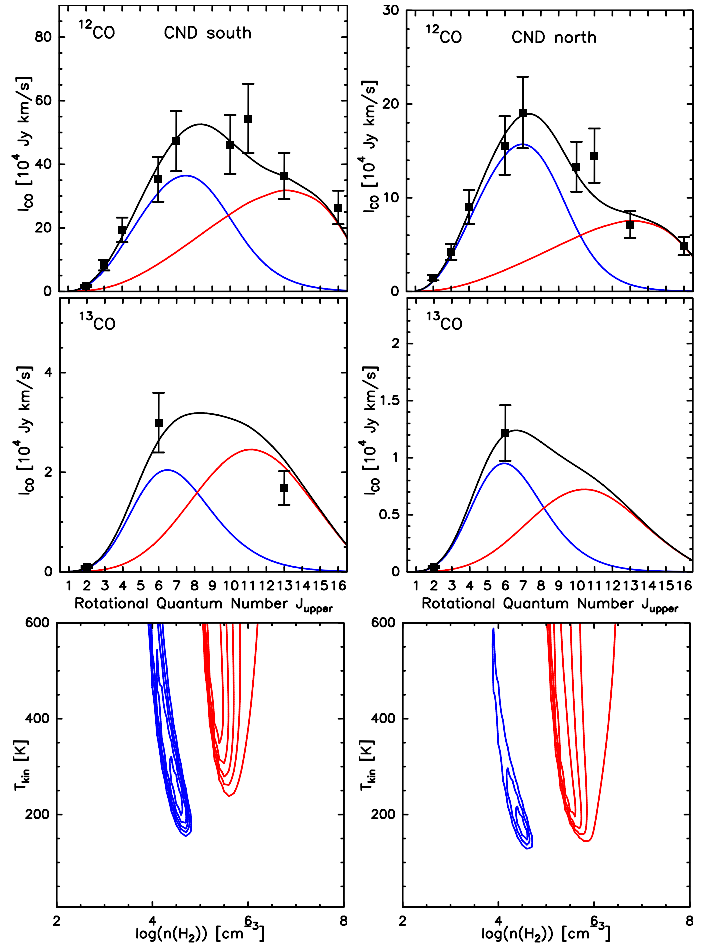}
\caption{LVG model fits with the parameters given in Table 2 for CND-S in the left column, for CND-N in the right column. The lowest panels display the $\chi^2$ distribution of allowed solutions in the density-temperature space (units are from 4 to 14 by 2).}
\label{cospec2}
\vspace{-0.3cm}
\end{figure}

Fig.~2 compares the high-J $^{12}$CO spectra observed with GREAT with
those of lower rotational transitions. At negative velocities all
low-rotational lines are strongly affected by deep line-of-sight
absorption features; these are negligible for J$_{\rm{up}}$ $> 6$
(E$_{\rm{up}}$ $>$116 K). The FIR lines (J$_{\rm{up}}$ $\geqslant$
10) show clean emission profiles from the CND only. To recover
the flux affected by these absorptions, we fitted a high-order polynomial
to the broad emission masking the absorptions, and used the fit to estimate the
underlying emission.
Similarly, line contributions from the massive GC clouds (at and below +50 km/s)
disappear for J$_{\rm{up}}$ $\geqslant$10. Comparison of the observed
isotopic line ratios ($\thicksim$10-20, Table 1) with the isotopic
abundance ratio determined for Galactic center clouds,
[$^{12}$C]/[$^{13}$C] $\thicksim$25 \citep{gues85, wil94, martin12}
indicates that the CO emission from the CND is
only marginally optically thick, with opacities on the order of unity.

The line profiles of the high-J CO lines show higher intensities for the most negative
velocities in the line wings of CND-S as the J increases.
The rotation curve of the CND shows that velocity rises toward SgrA$^*$
until just beyond the CND-S and CND-N positions, so the increased
velocity indicates that the higher-J lines are displaced toward the
sharp inner edge of the CND.  While this could indicate heating by the
central sources, the lack of a strong shift at CND-N suggests that
local heating is also important.  Combined with the broader lines toward the brighter
CND-S position, this suggests that shocks from turbulence within the
CND might provide substantial local heating; \citet{chris05} also reported
 concentrations of  dense HCN gas cores associated with 2.2\,$\mu$m emission from
shock-excited H$_2$ at these positions
\citep[see, ][]{yusef01}.  Shocks will allow material to drop into the central cavity,
while radiative excitation will drive material out of the cavity.

Velocity-integrated brightness temperatures, corrected for
line-of-sight absorption in multi-component Gaussian fits, are compiled in
Table 1. The integration is limited to velocity ranges characteristic
for the southern [-150 to -30 km/s] and northern [80 to 150 km/s]
lobe, respectively, thereby excluding velocities with heavy
line-of-sight absorption and contributions from the massive GC
clouds.
In Fig.~2 the corresponding CO line spectral energy
distributions (SED) are displayed.  The peaks of the SEDs
at J ${\footnotesize{\gtrsim}}$ 7, 155 K above ground
reflect the existence of gas at high temperatures, and the width of
the distribution requires contributions from more than a single gas
phase.  Assuming local thermal equilibrium (LTE), we can estimate the optical depth ($\tau_{6-5}
\sim$ 1.8, Table 1) and the excitation temperature of the bulk
emission from the $^{12}$CO
and the $^{13}$CO(6-5) lines . Toward CND-S (clump "Q" of
\citet{mon09} and "O" of \citet{chris05}, with size of
$\sim$8$''$) the main beam temperature T$_{mb}$ $\sim$50 K observed in
the original 8.9$''$ APEX beam translates into an R-J corrected clump
excitation temperature of $\sim$135 K.  This provides a lower limit
for the kinetic temperature in this clump.

The CO excitation was investigated using a large velocity gradient
(LVG) model in a similar manner as described in \citet{weiss07}.  The
CO abundances in the GC region are well constrained by
[$^{12}$CO]/[H$_2$] = $8\cdot10^{-5}$ \citep[see ][for
details]{brad05} and, as commented before, the carbon isotopic ratio
[$^{12}$CO]/[$^{13}$CO] = 25.  Furthermore, the velocity gradient
($dv/dr$) in the inner CND can be estimated from clump sizes and line
widths observed in high spatial resolution HCN and CS interferometry,
which yields $dv/dr=150$ km\,s$^{-1}$\,pc$^{-1}$
\citep{chris05,mon09}. Then our model only has three free parameters
for each gas component, namely the H$_2$ density, the kinetic
temperature T$_{kin}$, and the effective source solid angle (filling
factor) expressed as equivalence clump radius r$_0$. In this reduced
parameter space the line opacities, which are most critical to fit the
observed $^{12}$CO to $^{13}$CO line ratios, are directly proportional
to the H$_2$ density ($\tau \propto N(\rm {CO}/dv) \propto
$[CO]/[H$_2$]\,n(H$_2$)\,$(dv/dr)^{-1}$).  As such our models have
high constraining power on the gas density while T$_{kin}$ and r$_0$
are partly degenerated.

The best-fit LVG model parameters to the SEDs are presented in Table 2
and displayed in Fig.~2. The main conclusion toward the
two selected bright rim positions are that
\begin{itemize}
\item 
the broad, structured SEDs cannot be fitted by a single gas
  component. We must describe the distribution by at least a
  two-component fit as an approximation to a likely continuous
  temperature/density distribution.\\
  In our two component approach the high-J ($>$10) transitions are
  entirely driven by the high excitation component.
  This component is more pronounced in the south than in the north position.
 Our models suggest that the difference is mainly due to a larger source size of
 this component in the southern lobe (Table 2). The physical parameters of the
two gas phases are similar between CND-N and CND-S.
  The 
  $^{12}$CO line temperatures (with increasing line brightness up to
  J=4-3, Table 1) reflect the lines' low optical depth, driven by
  the high turbulence. The LVG optical depths in the $^{12}$CO(6-5) lines
  are consistent with our LTE estimates above and predict optically
  thin ($\tau < 0.1$) emission for the two lowest CO transitions.

\item
Densities are tightly constrained in the reduced parameter space.
  We find H$_2$ gas densities of log(n) = 4.5$^{+0.2}_{-0.6}$ cm$^{-3}$ for the
low-excitation component for both regions. The gas density of the high-excitation component	
  in the southern lobe is log(n) = 5.2$^{+0.6}_{-0.2}$. Owing to the absence of
a $^{13}$CO(13-12) measurement toward CND-N, the parameters for its high-excitation gas
  are less well constrained. Values in Table 2 assume the same (low) opacity for
 this line as  derived for CND-S.   Higher gas densities are excluded because they would overestimate the line opacities.
 As such our models strongly suggest gravitational unbound clouds.
\item
 Molecular gas masses (for the restricted velocity ranges, Table~1) 
in the $22.5^{\prime\prime}$ beam toward CND-S are M$_{-150}^{-30}$ = 280 (105) M$_\odot$ for
  the lower (higher) excitation phase, and M$_{+80}^{+150}$ = 165 (40) M$_\odot$ toward
  CND-N. Assuming that the CO(7-6) emission at all velocities [-150,150] traces CND gas only (hence no contributions from l-o-s gas) the total gas masses toward these two positions are 795 and 590 M$_\odot$. Next, assuming comparable excitation conditions throughout the
  inner CND (defined by an outer disk radius of 3.3 pc), we can estimate its total gas mass by scaling the masses
  derived in the two beams (1385
  M$_\odot$) with the ratio of the total (spatially) integrated flux
  of the CO 7-6 transition across the CND (Fig. 1) to the CO(7-6) flux
  observed in these two beams. We derive a mass of the inner CND of
 1.2 $10^{4}$ M$_\odot$ - this is very comparable to the mass
  obtained by \citet{gen85} and, more recently, by \citet{etx11}, but
  much lower than results based on dense gas tracers (see discussion).

\item
Kinetic temperatures for the low-excitation gas are on the order of 200\,K, and 300-600K for the high-excitation component.
The lower limit for
  T$_{kin}$ of 150 K in all components arises because that
  lower temperatures would require gas densities in excess of the limits
  discussed above. Kinetic gas temperatures as low as the dust temperatures
  of 24/45\,K derived from recent {\it Herschel} observations \citep{etx11}
  are inconsistent with our isotopic line ratios even if we consider models
  with a lower [$^{12}$CO]/[$^{13}$CO] abundance ratio of only 40. Our models
  therefore strongly support shock heating as the most
  important mechanism because shocks are the only mechanism to drive the
  kinetic temperature significantly above the dust temperature.
\end{itemize}

\begin{table}
\caption{Parameters of best-fit LVG models}	       % title of Table
\label{seddata}      % is used to refer this table in the text
\centering			     % used for centering table
\begin{tabular}{l l c c c c c c c}	 % centered columns (4 columns)
\hline\hline	\vspace{-0.25cm}\\	
 & gas phase & r$_0$ (pc)  &T$_{kin}$	&log n(H$_{2}$)&log N(H$_2$)\\
 \vspace{-0.25cm}\\ \hline \vspace{-0.25cm}\\
\vspace{0.1cm}CND-S   & low exc.   &0.31     &200$^{+300}_{-70}$     &4.5$^{+0.2}_{-0.5}$    &22.64\\
 \vspace{0.1cm}       & high exc.  & 0.08     &500$^{+100}_{-210}$    &5.2$^{+0.4}_{-0.2}$    &23.34\\
\vspace{-0.25cm}\\
\vspace{0.1cm}CND-N   & low exc.   &0.32      &175$^{+425}_{-45}$     &4.5$^{+0.3}_{-0.7}$    &22.35\\
\vspace{0.1cm}        & high exc.  &0.06      &325$^{+275}_{-165}$     &5.3$^{+0.6}_{-0.3}$    &23.15\\
\hline 		       % inserts single horizontal line
\hline 				  %inserts single line
\end{tabular}
\tablefoot{For each of the two gas phases we quote the kinetic temperature and H$_2$ density, the equivalent radius
(source solid angle) and the H$_2$ column density. The latter is calculated via
N(H$_2$) = 3.09\,10$^{18}$\,n(H$_2$)\,dv\,($\delta$v/$\delta$r)$^{-1}$. Recall that model fits are performed in restricted velocity windows only, Table~1.
Uncertainties describe the 3 dB $\chi^2$ fall-off, see Fig.~2.}
\end{table}

\section{Discussion}
The gas densities derived from our CO excitation analysis clearly
imply that the CO emitting gas cannot be gravitationally stable
against tidal shear because they are much below the Roche limit of 10$^7$
cm$^{-3}$. This comparison and because the
observed velocity gradient is much higher than motions expected for
virialized clouds at our densities strongly suggest that the CND
clouds are transient features. This is in contrast to conclusions
gained from studies of dense gas tracers such as HCN and HCO$^+$
\citep[e.g. ][]{chris05, mon09}. These studies derive much higher gas
densities ($>$10$^7$ cm$^{-3}$), albeit under the assumption of
virialization. As a consequence, these studies yield molecular
gas masses on the order of $10^6$ M$_\odot$ for the CND, roughly two orders
of magnitude above our estimate.  We note that these differences cannot 
be explained by the different critical densities between CO and
HCN/HCO$^+$ because our high-J CO transitions have values comparable to
those of the low-J HCN line and even exceed those of HCO$^+$.

To resolve the obvious discrepancy between these classical density
tracers, a more rigorous investigation of the HCN/HCO$^+$ excitation
will be performed - including those higher excitation submm and FIR
transitions that have not been considered so far. Complementary
$^{13}C$ isotopologues shall be observed to constrain the opacity of
the lines. Infrared pumping via the strong
14\,$\mu$m radiation field in the CND will be studied  In the 
excitation models \citep[see ][
for an evaluation of the effect]{sak10,weiss07}. Observationally one
should aim to constrain this excitation path, and hence the value of
HCN to trace dense gas in the CND, by searching for the $v_2$
vibrationally excited rotational transitions.\\
Based on our GREAT/SOFIA data we conclude that the (inner) CND is
best described by a collection of transient filamentary streamers and
clumps \citep{gues1987}. Its mass is comparatively low, few 10$^4$
M$_\odot$, which has implications on the mass accretion rate toward
the central object \citep[][]{genzel2010}.

\begin{acknowledgements}
%      Part of this work was supported by the German \emph{Deut\-sche For\-schungs\-ge\-mein\-schaft, DFG\/} project number Ts~17/2--1.
      We thank the SOFIA engineering and operations teams whose support has been essential for the GREAT accomplishments during Early Science.
      Based [in part] on observations made with the NASA/DLR Stratospheric Observatory for Infrared Astronomy. SOFIA Science Mission Operations are conducted jointly by the Universities Space Research Association, Inc., under NASA contract NAS2-97001, and the Deutsches SOFIA Institut under DLR contract 50 OK 0901.
      \end{acknowledgements}

%\bibliography{CNDCOSED}%

\end{document}